\documentclass[useAMS,usenatbib]{mn2e}
\usepackage{graphicx}
\usepackage{txfonts}
\usepackage{amssymb}
\usepackage{amsfonts}
\usepackage{times}
\usepackage{hyperref}
\usepackage{mathbbol}

\pdfoutput=1

 \title[]
   {Biased cosmological parameter estimation with galaxy cluster counts in the presence of primordial non-Gaussianities}
\author[A. M. M. Trindade et al.]{A. M. M. Trindade$^{1,2}$\thanks{E-mail: Arlindo.Trindade@astro.up.pt}, P. P. Avelino$^{1,2}$ and  P. T. P. Viana$^{1,2}$ 
\\
$^1$ Centro de Asrof\'{\i}sica da Universidade do Porto, Rua das Estrelas 687, 4150-762 Porto, Portugal\\
$^2$ Departamento de F\'{\i}sica e Astronomia da Faculdade de Ci\^encias da Universidade do Porto, Rua do Campo Alegre 687, 4169-007 Porto, Portugal}

\begin{document}


\pagerange{\pageref{firstpage}--\pageref{lastpage}} \pubyear{2013}

\maketitle

\begin{abstract}
The redshift dependence of the abundance of galaxy clusters is very sensitive to the statistical properties of primordial density perturbations. It can thus be used to probe small deviations from Gaussian initial conditions. Such deviations constitute a very important signature of many inflationary scenarios, and are thus expected to provide crucial information on physical processes which took place in the very early Universe. 

We have determined the biases which may be introduced in the estimation of cosmological parameters by wrongly assuming the absence of primordial non-Gaussianities. Although we find that the estimation of the present-day dark energy density using cluster counts is not very sensitive to the non-Gaussian properties of the density field, we show that the biases can be considerably larger in the estimation of the dark energy equation of state parameter $w$ and of the amplitude of the primordial density perturbations.

Our results suggest that a significant level of non-Gaussianity at cluster scales may be able to reconcile
the constraint on the amplitude of the primordial perturbations obtained using galaxy cluster
number counts from the Planck Sunyaev-Zeldovich Catalog with that obtained from the
primary Cosmic Microwave Background anisotropies measured by the Planck satellite. 

\end{abstract}
 

\section{Introduction}

Inflation is widely accepted as the most elegant mechanism to generate the density fluctuations which seeded the Large-Scale Structure (LSS) of the Universe we observe today. The simplest standard, single field and slow-roll, inflation model predicts, among other things, that the primordial density perturbations were nearly Gaussian distributed (see e.g. \citealt{2003JCAP...10..003C,2003JHEP...05..013M,2005PhRvL..95l1302L,2005JCAP...06..003S,2007PhRvD..76h3004S}). While such predictions seem to be in good agreement with current observations of the cosmic microwave background (CMB) anisotropies (e.g. \citealt{2008JCAP...08..031S}) and large-scale structure (e.g. \citealt{2011ApJS..192...18K}), a significant, potentially observable level of non-Gaussianity may be produced in many inflationary models. Detecting and constraining primordial non-Gaussianities has become a crucial task in current cosmological studies, since a positive detection of non-Gaussianity  would rule out a considerable number of inflationary models, opening an entirely new window into the very early Universe. 

A wide range of observables have been used to probe primordial non-Gaussianities. While the three-point statistics of the temperature anisotropies in the CMB is the most common tool, the statistical properties of the large-scale structure, namely the bispectrum and/or trispectrum of the galaxy distribution (e.g. \citealt{2007PhRvD..76h3004S,2008ApJ...677L..77M},\citealt{2012MNRAS.422.2854G}) and weak-lensing observations (e.g. \citealt{2012MNRAS.421..797S,2012MNRAS.426.2870H}), as well as CMB-LSS \citep{2012arXiv1205.0563T} and CMB-21cm \citep{2012PhRvD..85d3518T} cross-correlations have been used to the same effect. The evolution with time of the abundance of massive collapsed objects such as galaxy clusters (see e.g. \citealt{2000ApJ...541...10M,2000MNRAS.311..781R}) also holds key information that could be used to probe primordial non-Gaussianities. In fact, we have shown \citep{2012MNRAS.424.1442T} that assuming the absence of primordial non-Gaussianities, may lead to an apparent discontinuity in the evolution of the estimated effective equation of state parameter with redshift using galaxy cluster counts. In the same spirit, we will estimate and quantify the biases which might be introduced in the determination of several of the most important cosmological parameters using the evolution with time of the galaxy cluster abundance, if it was wrongly assumed that the initial conditions of the primordial density field were Gaussian distributed. We focus our attention on the present-day dark energy density, $\Omega_{de}$, the dark energy equation of state parameter $w$ (here assumed to be a constant), and the present-day root mean square mass perturbations, $\sigma_{8}$, at the standard $8 \, h^{-1} \, {\rm Mpc}$ scale. Primordial non-Gaussianity is usually parametrized by a non-linear parameter, $f_{NL}$, which relates the bispectrum and power spectrum of the primordial curvature perturbation \citep{2008JCAP...04..014L,2012MNRAS.424.1442T}. The parameter $f_{NL}$ can be calculated using different prescriptions, each associated with a different physical mechanism for the generation of primordial non-Gaussianities. We will mostly consider the so-called Local configuration (see \citealt{2012MNRAS.424.1442T} for more details), although we will also mention at some point the Equilateral configuration. 


\section{Galaxy Cluster Abundance}

The abundance of galaxy clusters, as a function of mass and redshift, contains information which can be used to compare cosmological models or constrain the value of some parameters associated with them. This observable is sensitive both to the volume of space and to the growth of structure on scales of the order of a few tens of Mpc, as a function of redshift, more specifically, to the expansion history of the Universe and the amplitude of the primordial density perturbations on those scales. However, even slight deviations from Gaussianity in the probability distribution of the primordial density perturbations can have a measurable impact on the galaxy cluster abundance, especially at hight redshift and masses 
(see \citealt{2000ApJ...541...10M}, \citealt{2007ApJ...658..669S}, and references therein).

The number of galaxy clusters in bins of redshift, centered at redshift $z_{i}$ and with width $\Delta z$, can be computed as
\begin{equation}
\label{Nzbin}
 N\left(z_{i}\pm \Delta z \right)=f_{sky}\int_{z_{i}-\Delta z/2}^{z_{i}+\Delta z/2} \frac{dV}{dz} \left[\int_{M_{lim}\left(z\right)}^{\infty} dM\frac{dn}{dM} \right]\,,
\end{equation}
where $f_ {sky}$ is the fraction of sky being observed, $dn/dM$ is the halo mass function and $dV/dz$ is the comoving volume element, which in the case of a flat cosmology, is given by
\begin{equation}
  \label{volume}
\frac{dV(z)}{dz}=4\pi\, \chi\left(z\right) \frac{d \chi(z)}{dz}\,,
 \end{equation}
with $\chi \left(z\right)$ being the comoving radial distance.

We will use the same prescription as in \cite{2012MNRAS.424.1442T} to calculate the halo mass function in the presence of non-Gaussianities in the probability distribution of the primordial density perturbations, which in turn followed \cite{2008JCAP...04..014L}, 

\begin{equation}
\label{mass_function_new}
\frac{dn_{NG}}{dM}\left(z,M,f_{NL}\right)=\frac{dn_{ST}}{dM}\frac{dn_{PS}/dM(z,M,f_{NL})}{dn_{PS}/dM(z,M,f_{NL}=0)}\,.
\end{equation}
where
\begin{eqnarray}
\label{ST_mass_function}
\frac{dn_{ST}}{dM}\left(z,M\right)&=&-\sqrt{\frac{2a}{\pi}}A\left(1+\left(\frac{a\delta_{c}}{\sigma^{2}}\right)^{-p}\right)
\frac{\bar{\rho}}{M^{2}}\frac{\delta_{c}\left(z\right)}{\sigma_{M}} \nonumber \\
 &\times& \frac{d\ln\,\sigma_{M}}{d\ln\, M} \mathit{e}^{-a\delta_{c}^{2}\left(z\right)/\left(2\sigma_{M}^{2}\right)}\,,
\end{eqnarray}
with $a=0.707$, $A=0.322184$ and $p=0.3$, is the Seth-Tormen mass function \citep{2001MNRAS.323....1S}, and
\begin{eqnarray}
\label{ng_mass_function}
\frac{dn_{PS}}{dM}\left(M,z\right)=-\sqrt{\frac{2}{\pi}}\frac{\overline{\rho}}{M^{2}}\mathit{e}^{-\delta_{c}^{2}\left(z\right)/2\sigma_{M}^{2}}\left[\frac{d\ln \sigma_{M}}{d\ln M}
\left(\frac{\delta_{c}\left(z\right)}{\sigma_{M}} + \frac{S_{3M}\sigma_{M}}{6} \right. \right. \nonumber \\ 
\left. \left. \times \left(\frac{\delta_{c}^{4}\left(z\right)}{\sigma_{M}^{4}} -2\frac{\delta_{c}^{2}\left(z\right)}{\sigma_{M}^{2}}-1 
 \right) \right) +\frac{1}{6}\frac{dS_{3M}}{d\ln M}\sigma_{M}\left(\frac{\delta_{c}^{2}\left(z\right)}{\sigma_{M}^{2}}-1\right)\right]\,,
\end{eqnarray}
where $S_{3M}$ is the skewness of the smoothed density field [if $f_{NL}=0$, then $S_{3M}=0$ and Eq. (\ref{ng_mass_function}) reduces to the Gaussian halo mass function]. This later formula comes from an extension to non-Gaussian density fields of the Press - Schechter formalism \citep{1974ApJ...187..425P}, proposed by \cite{2008JCAP...04..014L}. 
It has been shown to provide a good fit to results from N-body simulations (see \citealt{2010JCAP...10..022W} and references therein). 

The present-day value of the variance of the density perturbations, $\delta_{R}$, smoothed on a scale $R$ can be computed as, 
\begin{equation}
\label{2point_function}
\sigma^{2}\left(R\right)=\delta_{R}^{2} =\int \frac{d^{3}\mathbf{k_{1}}}{\left(2\pi\right)^3}\int \frac{d^{3}\mathbf{k_{2}}}{\left(2\pi\right)^3}
\mathcal{F}_{1} \mathcal{F}_{2} \langle \zeta_{1} \zeta_{2}\rangle,
\end{equation}
while the three-point function for the smoothed density field is given by \citep{2008JCAP...04..014L}
\begin{eqnarray}
 \label{3point_corr}
 \langle \delta_{R}^{3}\rangle =f_{NL} \int \frac{d^{3}\mathbf{k_{1}}}{\left(2\pi\right)^{3}}\int \frac{d^{3}\mathbf{k_{2}}}{\left(2\pi\right)^{3}} 
\int \frac{d^{3}\mathbf{k_{2}}}{\left(2\pi\right)^{3}}
\mathcal{F}_{1} \mathcal{F}_{2} \mathcal{F}_{3}\langle \zeta_{1} \zeta_{2} \zeta_{3}\rangle,
\end{eqnarray}
with $\zeta_{i}\equiv \zeta\left(\mathbf{k_{i}}\right)$, $\mathcal{F}_{i}\equiv W\left(k_{i},R\right)\mathcal{M}\left( k_{i}\right)T\left(k_{i}\right)$, where $T\left(k\right)$ is the transfer function adopted from \cite{1986ApJ...304...15B}, $M\left(k\right)=\left(2/5\right)c^{2} \Omega_{m}^{-1} H_{0}^{-2} k^{2} $, $W\left(k,R\right)$ is the smoothing top-hat window and we use the shape parameter given by \cite{1995ApJS..100..281S}, ${\Gamma=\Omega_{m}h\exp \left[-\Omega_{b}\left(1+\sqrt{2h}/\Omega_{m}\right)\right]}$.

\begin{figure*}
\centering

\includegraphics[scale=0.28]{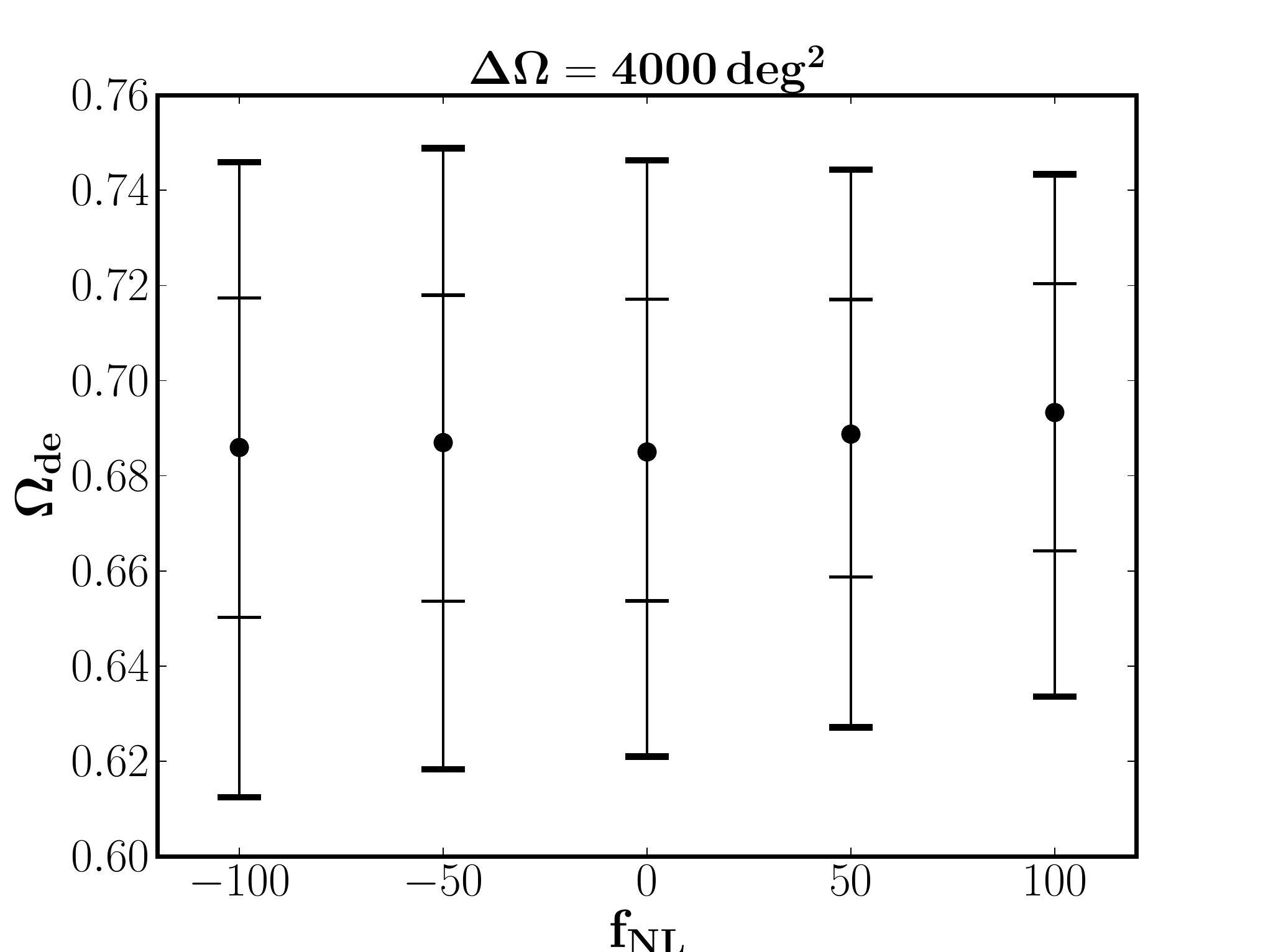}
\includegraphics[scale=0.28]{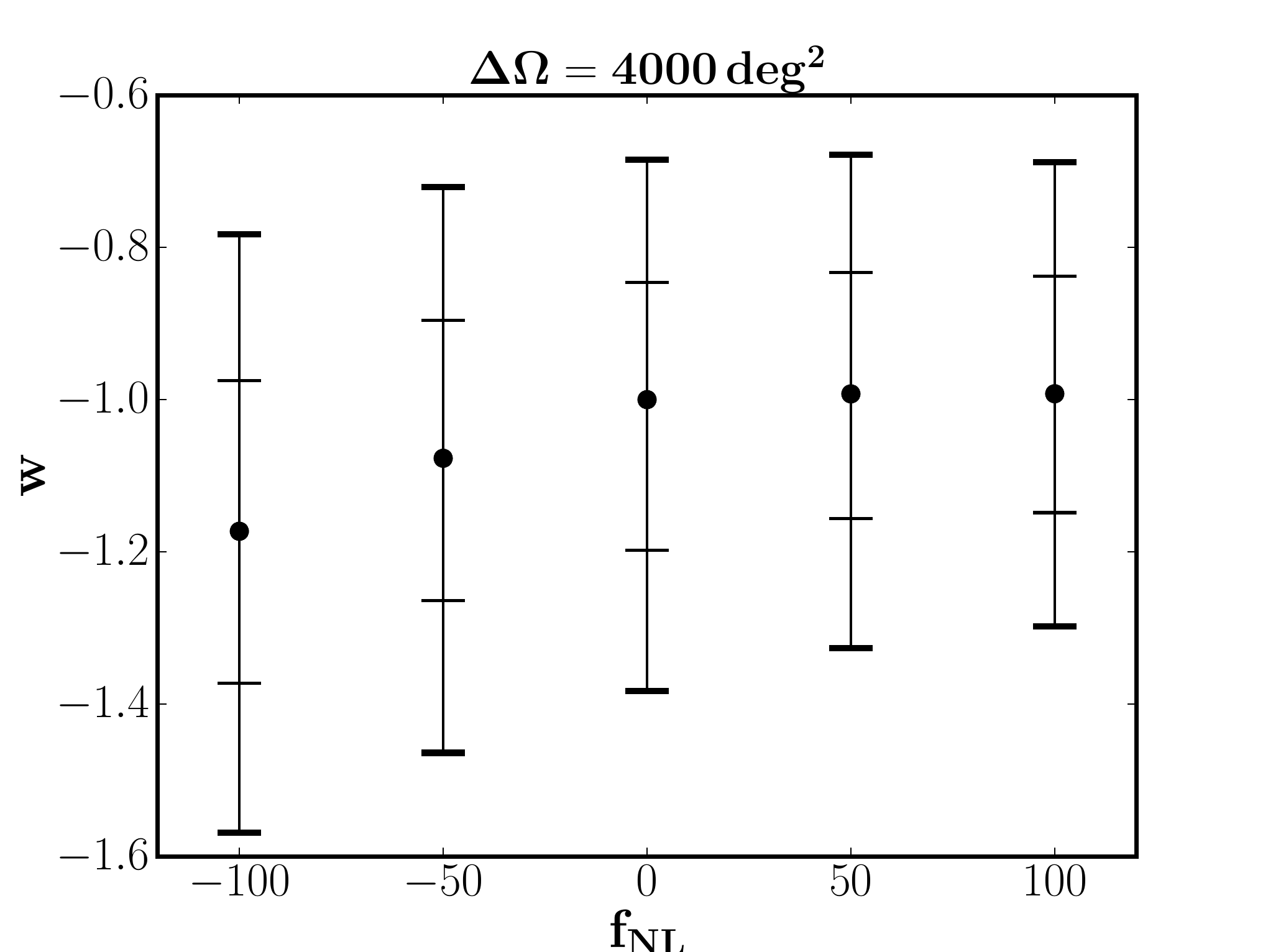}
\includegraphics[scale=0.28]{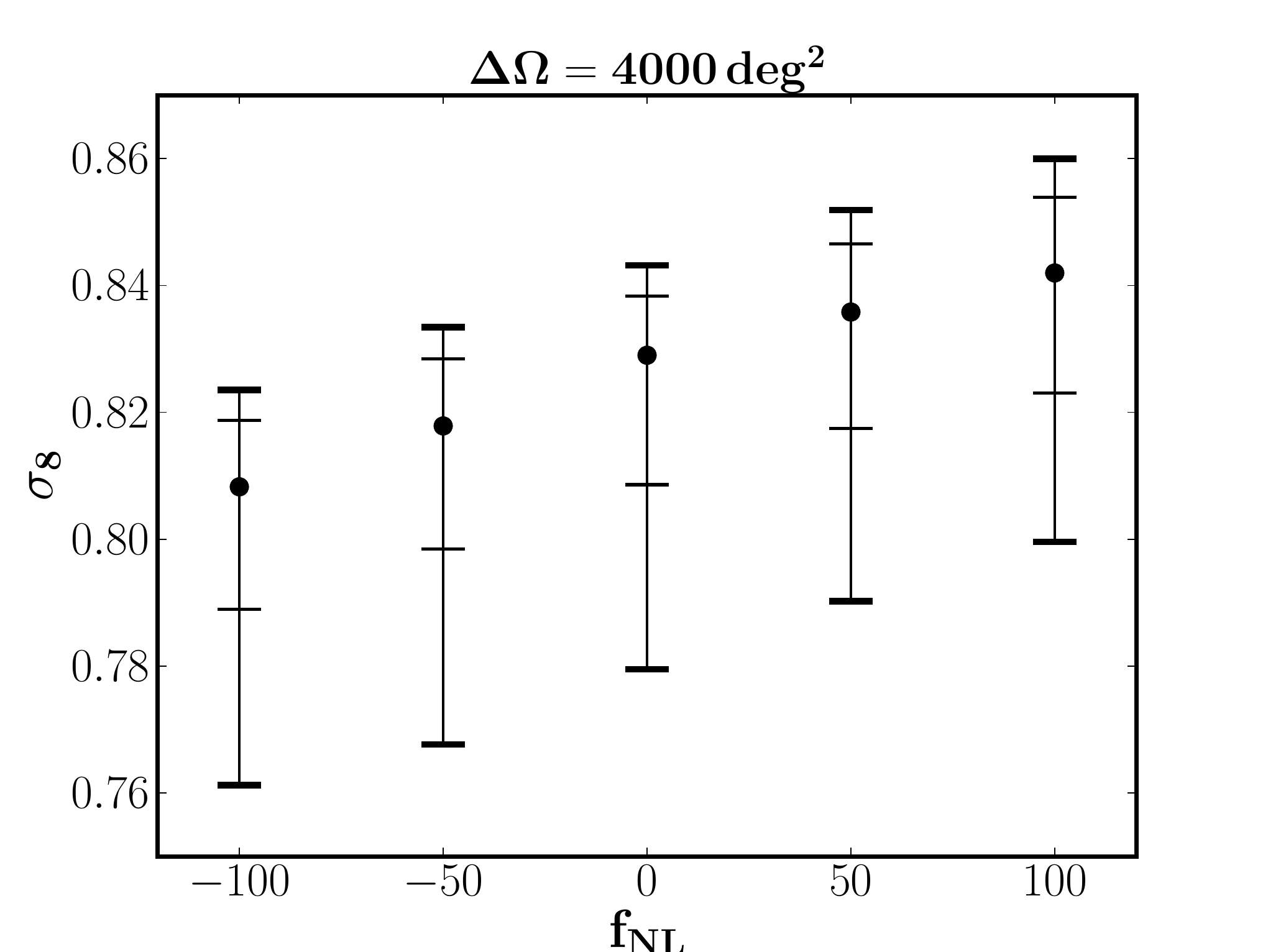}

\label{Figure1}
\caption{The most probable values of $\Omega_{de}$, $w$ and $\sigma_{8}$, and associated  $1\sigma$ (inner thin ticks) and $2\sigma$ (outer thick ticks) confidence levels, for a sky coverage of $4000\, \rm deg^ 2$.}
\end{figure*}
\section{Results}

In order to quantify the biases which may arise in the estimation of cosmological parameters using the redshift evolution of the galaxy cluster abundance due to wrongly assuming the absence of primordial non-gaussianities, we will follow the same approach considered in \cite{2009MNRAS.397..577S}. Therefore, we will assume that the likelihood, $\mathcal{L}$, of observing a given number of clusters, for a certain combination of cosmological parameters, in each bin of redshift and mass is given by
\begin{equation}
\label{likelihood}
\ln \mathcal{L}=\sum_{i=1}^{n_{z}}\sum_{j=1}^{n_{M}} \left[ N_{ij}^{m}\ln N_{ij}^{f}-N_{ij}^{f} -\ln \Gamma \left(N_{ij}^{f} +1 \right)\right] \,,
\end{equation}
where $n_z$ and $n_{M}$ are respectively the number of redshift and mass bins, while $N_{ij}^{f}$ and $N_{ij}^{m}$ are the number of counts for respectively the fiducial combination of cosmological parameters and the combination assumed to be true, for the $i$-th redshift bin and the $j$-th  mass bin. Thus, our assumed observed cluster catalogue will not be a particular realization of the cluster redshift distribution in the context of the fiducial model, but a sort of average-catalogue. Nevertheless, this will allow for a very good estimate of the size and shape of the expected likelihood contours expected from an analysis of the information contained in a real cluster catalogue, and avoid the offset in the best fit away from the fiducial values for the cosmological parameters, that would result from using a randomly-generated cluster catalogue based on the fiducial model (for more details see \citealt{2009MNRAS.397..577S}).

In the above expression for the likelihood, we assumed that all clusters are randomly distributed in space, i.e. their positions follow a Poisson distribution. However, those positions are in fact spatially correlated. The effect of these correlations on the cluster abundance is often referred to as cosmic variance, while the variations in the cluster abundance due to the cluster distribution being a Poisson process are known as shot-noise. It has been shown, assuming primordial Gaussian density perturbations, that the contribution of the cosmic variance to the statistical uncertainty associated with the galaxy cluster abundance is at least an order of magnitude smaller than the contribution due to shot-noise, almost independently of the surveyed sky area, as long as the cluster mass threshold is above $5\times10^{14}h^{-1}M_{\odot}$ \citep{2011A&A...536..A95V}. Therefore, so that we can safely neglect the contribution from cosmic variance, we will set the cluster mass threshold to $5\times10^{14}h^{-1}M_{\odot}$, noting that the level of non-Gaussianity we are assuming is relatively small, not affecting much the galaxy cluster abundance with respect to the Gaussian case \citep{2012MNRAS.424.1442T}. 

We assume a flat $\Lambda$CDM fiducial cosmology (i.e. the fiducial value for $w$ is assumed to be $-1$), as derived in \cite{2013arXiv1303.5076P}, namely, a Hubble constant, $H_{0}$, equal to $100h\,{\rm km}\,{\rm s}^{-1}\,{\rm Mpc}^{-1}$ with $h=0.673$, fractional energy densities associated with dark energy and baryons today of $\Omega_{de}=0.685$ and $\Omega_{b}h^{2}=0.02205$ respectively, a scalar spectral index, $n_{s}$, equal to $0.9603$, and  power spectrum normalization of $\sigma_{8}=0.829$. The level of non-gaussianity is parametrized by the $f_{NL}$ parameter, which we allow to vary from -100 up to 100 with increments of 50. We also consider a fiducial sky area of $4000\, {\rm deg}^{2}$ and we generate the expected number of clusters in redshift bins with width $\Delta z=0.1$ up to a redshift of 1.4. The cosmological parameters whose values we attempted to constrain using the information contained on those catalogues were $\Omega_{de}$, $w$ and $\sigma_{8}$, where $w$ is the (assumed independent of redshift) constant equation of state associated with the dark energy. Flat priors were associated with all, rendering their posterior probabilities proportional to the likelihood given by Eq. \ref{likelihood}. We set $f_{NL}=0$ as a prior, given our objective of quantifying the biases which may arise due to wrongly assuming the absence of primordial non-gaussianities. The exploration of the likelihood in the defined parameter space was carried out using a custom code based on standard Monte Carlo Markov Chain techniques (e.g. \citealt{2011RevModPhys.83.943}).

As can be seen in Fig. 1, the biases on $\Omega_{de}$ and $w$, that arise from wrongly assuming $f_{NL}$ to be zero, are very small. In fact, it would be necessary larger values of $|f_{NL}|$ than the ones considered here, combined with multiple mass bins and a larger sky coverage for the fiducial values of  $\Omega_{de}$ and/or $w$ to fall outside the derived $2\sigma$ confidence levels. 

Contrary to $\Omega_{de}$ and $w$, the bias on the estimation of $\sigma_{8}$ due to wrongly assuming gaussianity is significantly more severe, although even in this case the exclusion of the fiducial value at more than $2\sigma$ requires a value for $f_{NL}$ lower than about -80, or significantly higher than 100. The same exclusion level is attained for lower $|f_{NL}|$ if multiple mass bins are considered and/or the sky area coverage increased. In the later case, we have confirmed that as expected, the uncertainty associated with the estimation of each parameter is inversely proportional to the square root of the sky area coverage. For example, increasing it to $40000\, deg^2$ would result in the fiducial value for $\sigma_8$ assumed here to be excluded at more than $2\sigma$ for $f_{NL}$ smaller than $-30$. 

The best fit values of $\Omega_{de}$, $w$ and $\sigma_{8}$ as a function of the non-Gaussianity level are given approximately by
\begin{eqnarray}
\label{fits1}
\Omega_{de}&=& 0.686+3.3\times 10^{-5}f_{NL}+3.7\times 10^{-7}\left(f_{NL}\right)^2\\
w&=&-1.010+8.9\times 10^{-4} f_{NL}-7.4\times 10^{-6}\left(f_{NL}\right)^2 \\
\sigma_{8}&=& 0.829+1.7\times 10^{-4} f_{NL}-3.2\times 10^{-7}\left(f_{NL}\right)^2,
\end{eqnarray}
with a maximum fitting error below $1\%$. The induced systematic errors in the cosmological parameters due to a systematic error in the non-Gaussianity parameter can be obtained by simply differentiating Eqs. 9 to 11 with respect to $f_{NL}$. Although the results above are specifically 
for the local parametrization of $f_{NL}$, they can be used to infer the what would happen had we considered the equilateral parametrization, given that a value for $f^{equilateral}_{NL}$ which is 3 times some $f^{local}_{NL}$ will yield roughly the same cluster abundance. We have also found that the dependences expressed through Eqs. 9 to 11 do not change significantly if the cluster mass threshold, assumed here to be $5\times10^{14}h^{-1}M_{\odot}$, is changed.

We have found that the abundance of galaxy clusters can also be computed for the equilateral parametrization using
\begin{equation}
\label{rescaling}
f_{NL}^{local}\simeq\frac{f_{NL}^ {equil}}{3.6}\,.
\end{equation}
which seems to be consistent with the recent results of \cite{2013arXiv1304.1216S} and do not change much with cluster mass threshold.
\vspace{0.23cm}
\section{Conclusions}

In this paper we have computed the dependence, as a function of $f_{NL}$, of the biases that arise in the estimation of the
cosmological parameters $\Omega_{de}$, $w$ and $\sigma_8$, when it is (wrongly) assumed, for the purpose of the
statistical analysis, that the density field is Gaussian distributed. We have found that such biases are quite small for the 
first two parameters, but significant in the case of $\sigma_8$, in particular in the face of the high statistical accuracy with 
which this parameter is expected to be determined in the near-future (e.g. \citealt{2012MNRAS.422.44P}).

If $f_{NL}$ is assumed to be scale-independent, then the recent results obtained by the Planck team \citep{2013arXiv1303.5084P}, impose
severe constraints on the amount of non-Gaussianity at cluster scales ($f_{NL}=2.7 \pm 5.8$
for the local parameterization) making it safe to neglect primordial non-Gaussianities in the
determination of cosmological parameters using the galaxy cluster abundance. However, $f_{NL}$ could be a function of scale 
(see \citealt{2008JCAP...04..014L} and references therein) , in which case the effects of non-gaussianity on the cluster abundance may need to be taken into account in 
the determination of $\sigma_8$. In fact, Eq. (11) suggests that a significantly negative value for  $f_{NL}$ at cluster scales, of the order of $-240$, 
would reconcile the constraint on $\sigma_8$ obtained using the cluster abundance inferred from the Planck 
Sunyaev-Zeldovich Catalogue ($\sigma_8=0.77 \pm 0.02$, \citealt{2013arXiv1303.5080P}) with the one obtained from the properties of the 
primordial CMB temperature anisotropies ($\sigma_8=0.829 \pm 0.012$, \citealt{2013arXiv1303.5076P}).

\section*{Acknowledgements}

We thank Lara Sousa, Ant\'onio da Silva and Michael Bazot for useful discussions during the preparation of this paper. AMMT was supported by the FCT/IDPASC grant contract SFRH/BD/51647/2011. AMMT, PPA and PTPV acknowledge financial support from project 
PTDC/FIS/111725/2009, funded by Funda\c{c}\~{a}o para a Ci\^{e}ncia e a Tecnologia. 
{\small
\bibliographystyle{aa}

}

\end{document}